# Softening the ultra-stiff: controlled variation of Young's modulus in single-crystal diamond

A. Battiato[1,2,3,4], M. Lorusso[5], E. Bernardi[1,2,3,4], F. Picollo[3,1,2,4], F. Bosia[1,2,3,4]*,

D. Ugues[6], A. Zelferino[1], A. Damin[7,2], J. Baima[7,2], N. M. Pugno[8,9,10], E. P. Ambrosio[5],

P. Olivero[1,2,3,4]

[1] *Physics Department, University of Torino, Torino, Italy*

[2] *Inter-departmental Centre "Nanostructured Interfaces and Surfaces" (NIS), University of Torino, Torino, Italy*

[3] *INFN - National Institute of Nuclear Physics, section of Torino, Torino, Italy*

[4] *CNISM - Consorzio Nazionale Interuniversitario per le Scienze fisiche della Materia, section of Torino, Torino, Italy*

[5] *Center for Space Human Robotics @ Polito, Istituto Italiano di Tecnologia, Torino, Italy*

[6] *Department of Applied Science and Technologies, Politecnico of Torino, Italy*

[7] *Chemistry Department, University of Torino, Torino, Italy*

[8] *Department of Civil, Environmental and Mechanical Engineering, University of Trento, Italy*

[9] *Center for Materials and Microsystems, Fondazione Bruno Kessler, Trento, Italy*

[10] *School of Engineering and Materials Science, Queen Mary University of London, UK*

*\* Corresponding Author: federico.bosia@unito.it*

## Abstract

A combined experimental and numerical study on the variation of the elastic properties of defective single-crystal diamond is presented for the first time, by comparing nano-indentation measurements on MeV-ion-implanted samples with multi-scale modeling consisting of both *ab initio* atomistic calculations and meso-scale Finite Element Method (FEM) simulations. It is found that by locally introducing defects in the $2\times10^{18}$ - $5\times10^{21}$ cm$^{-3}$ density range, a significant reduction of Young's modulus, as well as of density, can be induced in the diamond crystal structure without incurring in the graphitization of the material. *Ab initio* atomistic simulations confirm the experimental findings with a good degree of confidence. FEM simulations are further employed to verify the consistency of measured deformations with a stiffness reduction, and to derive strain and stress levels in the implanted region. Combining these experimental and numerical results, we also provide insight into the mechanism responsible for the depth dependence of the graphitization threshold in diamond. This work prospects the possibility of achieving accurate tunability of the mechanical properties of single-crystal diamond through defect engineering, with significant technological applications, i.e. the fabrication and control of the resonant frequency of diamond-based micromechanical resonators.

# 1. Introduction

Diamond is an extremely attractive material for a broad range of technological applications due to its unique physical and chemical properties. In particular with regards to its extreme mechanical and thermal properties, in the past years several works were focused on developing mechanical structures and resonators in diamond either by MeV ion implantation [1, 2] or by reactive ion etching [2-6] with the purpose of taking advantage of its high mechanical hardness, stiffness and thermal conductivity. Moreover, diamond hosts a wide variety of luminescent defect centres [7, 8] that can act as stable single photon emitters at room temperature or as optically addressable solid-state spin-qubits [9, 10]. A challenging goal in this field is to efficiently couple negatively charged nitrogen-vacancy centres to resonant mechanical structures [11-13]. For advanced applications in nano-opto-mechanical devices, the prospect of being able to modify and finely tune the mechanical properties of diamond is therefore particularly appealing. In the case of other "2-D" carbon-based materials (e.g. carbon nanotubes or graphene), the effect of structural defects on their macroscopic mechanical properties has been studied both experimentally [14, 15] and theoretically [16, 17], while a significant gap is present in the case of bulk diamond. Here, we report such a systematic study, showing that a controlled modulation of the Young's modulus of diamond can be effectively achieved by defect engineering through MeV ion irradiation. Ion beam lithography based on MeV [18-20] or keV ions [21, 22] has emerged in the past years as one of the most promising techniques used in the micro- and nano-machining of diamond for different applications. It is known that ion implantation induces structural modifications (at low damage densities, consisting primarily of vacancies and interstitials [23]) and local mass density variations in the diamond crystal, which in turn result in mechanical deformations, including surface swelling [24] and local stresses [25, 26]. A mixed analytical/numerical approach to estimate these stresses has been developed, providing good agreement with experimental measurements [27]. One important issue that remains to be adequately addressed, however, is the variation of the elastic properties of damaged diamond as a function of induced damage density. This is particularly relevant for the Young's modulus, which is expected to vary between that of pristine diamond (i.e. ~1135 GPa [28, 29]) and that of amorphous carbon (i.e. ~20 GPa in a fully amorphized phase [30]). Clearly, this large variation in the elastic properties is likely to strongly affect the modeling of implantation-induced stresses. Attempts have been made to experimentally derive the variation of elastic properties in irradiated diamond [26], but only indirect estimations with limited accuracy were obtained. This significant lack of experimental evidence is partly due to the extremely high Young's modulus value of the pristine material, which makes it difficult to probe its mechanical properties. A related open question is represented by the relatively high uncertainty found in the literature on the value of the so-called "amorphisation threshold" of diamond, i.e. the damage level (usually parameterized with a vacancy density, as calculated via the SRIM Monte Carlo code [31]) above which the diamond lattice is permanently amorphized, and subsequently graphitizes upon thermal annealing. It is often hypothesized that the large variability of this parameter is related to the depth of implantation [32], as well as to self-annealing effects [33], but so far no unequivocal evidence of these effects has been provided.

In this work we present the first systematic study of the controlled variation of the elastic properties of diamond as a function on induced structural damage. The experimental measurements of the Young's modulus of MeV-ion-implanted diamond are performed with the nano-indentation technique and results are complemented by numerical simulations at two different scales, i.e. *ab initio* atomistic calculations of defected diamond supercells and Finite Element Method (FEM) simulations of full-field deformations and stresses at the meso-scale. Besides allowing a new level of control in the fine-tuning of mechanical properties of diamond-based mechanical structures, our analysis also allows a novel interpretation for the depth dependence of the amorphization threshold based on rigorous continuum mechanics considerations.

## 2. Experimental

*2.1. Sample preparation*

The sample under investigation was an artificial HPHT type Ib single crystal diamond sample synthesized by Sumitomo (Japan), 3×3×0.3 mm$^3$ in size, cut along the (100) crystalline direction with four optically polished faces, i.e. two opposite large surfaces and two opposite lateral surfaces. These "mechanical grade" diamond samples typically contain various impurities (N, Fe, Ni, Co) at concentration levels of the order of ~10-100 ppm, which do not affect significantly their mechanical properties, as confirmed by the test measurement performed from undamaged regions. The sample was implanted at room temperature on one of its lateral polished faces across its edge with one of its large surfaces, as schematically shown in Fig. 1, using a 2 MeV H$^+$ ion microbeam at the INFN Legnaro National Laboratories. A rectangular area of approximately 100×200 μm$^2$ was raster-scanned to deliver a uniform implantation fluence of 1×10$^{17}$ cm$^{-2}$.

SRIM simulations were carried out using the SRIM 2012.03 Monte Carlo code [31] to estimate the linear damage profile $\lambda(x)$, expressed as the number of induced vacancies per incoming ion per unit depth $x$. The calculations were carried out in "Detailed calculation mode with full damage cascade" mode, by setting the atom displacement energy value to 50 eV [34]. According to simulations, the irradiation conditions generate a strongly inhomogeneous damage density profile peaked at a depth of ~25 μm from the surface (see the inset of Fig. 1).

*2.2. Raman characterization*

Micro-Raman spectroscopy was employed to assess the degree of amorphization/graphitization in the regions of the diamond sample which were characterized by the highest ion-induced damage density, i.e. in correspondence with the end-of-range Bragg peak. The measurements were performed in the same cross-sectional geometry adopted for the nanoindentation measurements, i.e. the probing laser beam was focused across the upper surface of the sample and scanned across the ion-beam-induced damage profile (see Fig. 1 as a reference). A similar approach has been adopted in previous Raman studies in ion-implanted diamond [32, 35]. An inVia Raman micro-spectrometer (Renishaw) was employed, with a $\lambda = 514$ nm excitation laser beam focused onto a micrometric spot with a 80× objective. The laser power focused on the probed region was ~2.5 mW. A PC-controlled stage allowed the sample displacement across the three directions, and thus the mapping of the Raman signal with micrometric resolution. The experimental results are reported in Section 4.1.

*2.3. Nanoindentation measurements*

The lateral implantation geometry shown in Fig. 1 enabled to gain access to the upper surface of the sample with a nano-indentation profilometer, and thus a direct measurement of the structural and mechanical properties (i.e. hardness and Young's modulus) of the damaged diamond for a varying vacancy density profile. The choice of energetic light ions offered the advantage of inducing a relatively broad damage profile, so that the nanoindentation measurements were not severely influenced by edge effects. The nanoindentation measurements were carried out with a Hysitron TI 950 TriboIndenter, allowing high spatial resolution down to 500 nm and the probing of ultra-hard materials such as diamond thanks to the large applied loads (i.e. 5 mN). The instrument is equipped with *in situ* Scanning Probe Microscopy (SPM) imaging capability, using a Berkovich tip. The first step of the measurement consisted in the acquisition of a large-area SPM topography map, to determine the location of the Bragg peak and the corresponding swelling [36]. A typical Scanning Probe Microscopy (SPM) mapping of the measured region is shown in Fig. 2(a), highlighting a line along which a profile was acquired, from the sample edge to the bulk. The topography maps were acquired using the nano-indenter Berkovich tip in contact mode with a set-point of 2 μN a scan rate of 1 Hz and a scan size of 50×50 μm$^2$. A SPM profile from an unimplanted region near the indentation line was subtracted from the SPM curves, to eliminate instrumental effects related to sample tilt, thermal drift and nonlinearity of piezo-electric actuators. A 20×20 μm$^2$ map was then acquired in the proximity of the peak and nano-indentation was performed in the same region. The indentation profile was realized with 72 indents at a distance of 250 nm, resulting in a total profile length of 18 μm. A load of 5 mN was applied at a rate of 0.5 mN s$^{-1}$, applying the maximum load for each indentation for 2 s. In correspondence of every indent, "load vs displacement" curves were acquired, from which the elastic modulus of the material was derived using the approach proposed by Oliver and Pharr [37]. Indentation measurements were performed in the linear elastic range, both in unimplanted and in ion-implanted areas. As an example, Fig. 2(b) reports typical loading and unloading force-displacement curves collected from the unimplanted (dashed black curve) and implanted (solid blue curve) regions of the sample, with the latter measurement taken in correspondence with the implantation Bragg Peak. Measurements were performed with a set-point of 2 μN and a loading rate of 500 μN s$^{-1}$. Loading and unloading curves coincide, thus confirming that they were carried out in the linear elastic range. The measured reduced modulus $E_r$ is related to the Young's moduli and Poisson's ratios of both the indenter ($E_i$, $v_i$) and the sample ($E_s$, $v_s$), by the relation $1/E_r = (1-v_i^2)/E_i + (1-v_s^2)/E_s$ which is derived from contact mechanics [37], with $E_i = 1140$ GPa and $v_i = 0.07$. It is worth remarking that in this measurements both indenter and substrate were made of diamond, albeit with different defect densities, so that they contributed to the reduced modulus in similar proportions. In order to obtain a direct comparison between indentation and topography profiles, a SPM map of the region of interest was preliminarily acquired and four indentation measurements were performed at its corners. SEM analysis performed on the tip before and after the indentation revealed that no wear phenomena occurred during the tests. The experimental results are reported in Section 4.2.

## 3. Modelling

### 3.1. Ab initio simulations

An atomistic approach was employed to simulate the damage-induced modification of the Young's modulus in the material from first principles. To this purpose, the quantum-mechanical ab initio code CRYSTAL14 [38, 39] was employed, which allows the prediction of the structural and elastic properties of a defected material [40] through a "supercell" approach [41]. In this context, a supercell is defined as a multiple of the unit cell that contains a vacancy at its body center. The CRYSTAL software creates a bulk system by the repetition of such a defective supercell, so that (differently to the real physical system) the modeled system under investigation is homogeneous, i.e. defects are periodically distributed in the crystal. Supercells of different sizes allowed to simulate different defect densities. As shown in Fig. 3, single-defect cells with 128, 64, 32, 16 and 8 atoms were simulated, corresponding to vacancy densities of $1.4 \times 10^{21}$ cm$^{-3}$, $2.8 \times 10^{21}$ cm$^{-3}$, $5.5 \times 10^{21}$ cm$^{-3}$, $1.1 \times 10^{22}$ cm$^{-3}$ and $2.2 \times 10^{22}$ cm$^{-3}$, respectively. For each supercell, the geometry (both fractional coordinates and cell parameters) was preliminarily optimized. Then, the full elastic tensor was generated by deforming the unit cell. Second-order derivatives of the total energy with respect to the strain are evaluated in CRYSTAL as a numerical derivative of an analytical gradient [42, 43]. The Young's modulus was thus derived from the elastic constants tensor. Point symmetry was used at all stages of the calculation to reduce the number of components of the elastic tensor to be considered. For each irreducible strain, a deformation was applied to the system, so that $N$ strained configurations were defined according to a strain step. After a loop of $N$ strained configurations, the energy gradients were fitted with single-value-decomposition routines and the first derivatives were determined numerically. Hartee-Fock (HF) and B3LYP Hamiltonians were adopted [44, 45], both of which have been shown to produce particularly accurate results for this kind of system [44], and **the described procedure was adopted for both cases.** The neutral vacancy in diamond can occur in three spin states, i.e. $S_z$ = 0, 1, 2 [46, 47]. The spin state of the defect was however found to have a negligible influence on the elastic properties, so that the reported data only refer to the lowest-energy spin state $S_z$ = 0.

*3.2. Mesoscale model*

The dependence of structural/mechanical properties of damaged diamond have previously been estimated using a phenomenological mesoscale model accounting for damage saturation effects at high fluences, based on a "rule of mixtures" approach [27]. According to this model, when accounting for defect recombination in the crystal, the vacancy density $\rho_V$ in the depth direction $x$ can be expressed as:

$$\rho_V(x) = \alpha \cdot \left(1 - \exp\left[-\frac{F \cdot \lambda(x)}{\alpha}\right]\right) \qquad (1)$$

where $\lambda(x)$ is the linear vacancy density calculated numerically using the SRIM code, F is the implantation fluence and $\alpha$ is the saturation vacancy density. The latter $\alpha$ parameter can be derived as follows: let us consider an initial mass $M$ and volume $V_0$ of diamond that after implantation expands to a volume $V$ containing $n$ defects (i.e. vacancies), each assigned the same volume $v$ for sake of simplicity. The vacancy density is therefore $\rho_V = n/V$. By neglecting the mass of the implanted ions, the mass density of the implanted diamond volume $V$ is:

$$\rho = \frac{M}{V} = \frac{M}{V_0 + nv} \tag{2}$$

Since $\rho_d = M/V_0 = 3.52$ cm$^{-3}$ is the mass density of unimplanted diamond, we obtain:

$$\frac{1}{\rho} = \frac{V_0 + nv}{M} = \frac{1}{\rho_d} + \frac{nv}{M} = \frac{1}{\rho_d} + \frac{nv}{\rho V} = \frac{1}{\rho_d} + \frac{\rho_V v}{\rho} \tag{3}$$

By rearranging, we obtain:

$$\frac{\rho}{\rho_d} = 1 - \rho_V v = 1 - f \tag{4}$$

where $f = n \cdot v/V = v \cdot \rho_V$ is the volume fraction of vacancies, which correctly implies that $\rho = \rho_d$ for $f = 0$, and $\rho = 0$ for $f = 1$. In fact, in our case an upper bound exists for $f$ due to defect recombination effects [24] corresponding to a vacancy density saturation value $\rho_V = \alpha$. This vacancy density saturation value corresponds to an experimentally determined mass density saturation value of $\rho_{aC}$ = 2.06 g cm$^{-3}$ [48]. By rearranging Eq. (4) and estimating $v$ as the inverse of the atomic density of diamond ($\gamma = 1.77 \times 10^{23}$ cm$^{-3}$), we can derive $\alpha$ as:

$$\alpha = \gamma\left(1 - \frac{\rho_{aC}}{\rho_d}\right) \approx 7.3 \cdot 10^{23} cm^{-3} \tag{5}$$

consistently with previous works on 2 MeV H$^+$ implantations [49]. This vacancy density is shown in the inset of Fig. 1. The mass density variation $\rho(x)$ can thus be written as:

$$\rho(x) = \rho_d - (\rho_d - \rho_{aC}) \cdot \left(1 - \exp\left[-\frac{F \cdot \lambda(x)}{\alpha}\right]\right) \tag{6}$$

The Young's modulus dependence from the vacancy density can be derived from Quantized Fracture Mechanics (QFM) [17, 50] for the case of single isolated (non-interacting) vacancies:

$$E(x) = E_d \cdot \left(1 - \kappa \frac{\rho(x)}{\rho_d}\right) \tag{7}$$

where $\kappa$ is an empirical factor related to defect shape and interaction [17].

*3.3. Finite Element simulations*

Finite Element Model (FEM) simulations were carried out using the Structural Mechanics module of the COMSOL Multiphysics 5.0 package. A 3-D model of the implanted diamond sample was created and constrained expansion of the implanted diamond region due to local density reduction was simulated. The latter was numerically modelled according to elasticity theory by introducing residual strains $\varepsilon_i(x)$ in the three principal directions of the implanted material ($i = 1, 2, 3$):

$$\varepsilon_i(x) = \sqrt[3]{\frac{\rho_d}{\rho(x)}} - 1 \tag{8}$$

In this way, in an unconstrained expansion the volume variation would be inversely proportional to the density variation. The diamond sample was considered mechanically isotropic as a first approximation. Material density and Young's modulus spatial variations were accounted for using Eqs. (6) and (7). The same functional form, based on a rule of mixtures approach, was assumed for the variation of the Poisson's ratio:

$$\nu(x) = \nu_d - (\nu_d - \nu_{aC})\left(1 - \exp\left[-\frac{F \cdot \lambda(x)}{\alpha}\right]\right) \tag{9}$$

where $\nu_d = 0.07$ and $\nu_{aC} = 0.34$ are the Poisson's ratios of pristine diamond and amorphous carbon, respectively [51].

## 4. Results and discussion

*4.1. Raman measurements*

Figure 4(a) shows several Raman spectra acquired across the end-of-range depth, the blue plot corresponding to the end-of-range position (~25 μm). In the region characterized by the highest damage density a significant decrease of the first-order Raman line at 1332 cm$^{-1}$ is clearly observed which is associated to increasing structural damage, while no pronounced Raman features at ~1630 cm$^{-1}$ and 1680 cm$^{-1}$ (typically attributed to sp$^2$ defects [32, 52, 53]) can be observed. On the other hand, at frequencies lower than the 1$^{st}$ order transition, a broad band between 1000 cm$^{-1}$ and 1300 cm$^{-1}$ appears at the highest damage densities (blue plot), which is commonly attributed to sp$^3$ defects and in heavily stressed diamond [53, 54]. As for the features at frequencies above the first order transition, it is worth noting that the Raman peak at ~1370 cm$^{-1}$ closely resembles the D band at ~1350 cm$^{-1}$ commonly associated to tetrahedral amorphous carbon characterized by a high sp$^3$ content [55], while no G band at ~1580 cm$^{-1}$ is observed in our spectra. More likely, the ~1370 cm$^{-1}$ feature can be related (with a slight frequency shift attributed to the effect of different local stress fields) to a feature observed at ~1390 cm$^{-1}$ in highly stressed diamond regions close to laser-induced graphitized areas, which was attributed to sp$^3$ carbon allotropes (i.e. Z-carbon and hexagonal diamond) subjected to high mechanical stress [56]. A detailed analysis of the obtained Raman spectra is beyond the scope of this work, but it is nonetheless worth remarking that all of the above-mentioned attributions are compatible with a purely-sp$^3$ diamond-like phase, thus confirming that even in correspondence with the Bragg peak the damaged diamond structure is not subjected to a graphitization process. The progressive reduction of the first-order Raman transition is accompanied by its broadening and shift to lower frequencies. The latter feature is reported in Fig. 4(b), where the peak frequency is reported. The first-order Raman peak corresponding to the Bragg peak is positioned at ~1321 cm$^{-1}$. In [32], the redshift of the first-order Raman line in defective diamond has been correlated to vacancy density, estimated by the authors in a linear approximation as the product between the linear vacancy density predicted by SRIM and the implantation fluence. If such a correlation is applied to the measured ~1321 cm$^{-1}$ value, an estimated vacancy density at the Bragg peak of ~$4.5\times10^{21}$ cm$^{-3}$ is obtained in the above-mentioned linear approximation. If we consider that the SRIM-predicted linear vacancy density in such region is ~$4.5\times10^{4}$ cm$^{-1}$, we can obtain an estimation of the implantation fluence of ($4.5\times10^{21}$ cm$^{-3}$) / ($4.5\times10^{4}$ cm$^{-1}$) = $1\times10^{17}$ cm$^{-2}$, in striking agreement with the corresponding experimental value (see Section 2.1). Again, these correspondences confirm that the highly damaged regions of the sample can be still considered as a highly defective sp$^3$ diamond phase rather than a graphitized phase.

*4.2. Nanoindentation results*

Firstly, an unimplanted region of the sample was probed with the nano-indenter, yielding a value of $E_r = (600 \pm 45)$ GPa for the reduced Young's modulus, which is compatible with values from the literature [28, 29]. This measured reduced modulus value corresponds to an effective Young's modulus of $\hat{E} = E_s/[(1 - 2\nu_s)(1 + \nu_s)] = (1253 \pm 90)$ GPa, which can be adopted to express the relations between principal stresses $\sigma_{ii}$ and strains $\varepsilon_{ii}$ in the generalized 3D form of Hooke's law [57].

For the implanted diamond regions, we exploited the nanometric spatial resolution of the instrument to investigate the strongly inhomogeneous damage profile in implanted diamond in correspondence with the end-of-range Bragg peak (i.e. between 3 μm and 4 μm), where the variations in damage density and stiffness are more pronounced. Cross-sectional SPM topography and elastic modulus profiles of the implanted area, acquired from linear scans, are reported in Fig. 5. A ~10% decrease in the value of the reduced modulus is clearly visible in correspondence of the surface swelling peak. The position of the features along the depth direction match with the numerical predictions of SRIM Monte Carlo code (see the inset of Fig. 1).

*4.3. Density/stiffness variations and surface swelling results*

Experimental and numerical results for the relative variation of mass density and Young's modulus as a function of vacancy density are presented in Figs. 6a and 6b, respectively. For the mass density variation in Fig. 6a, the predictions of the model (see Eq. (6)) are compared to the results of *ab initio* HF and B3LYP simulations, showing good agreement and confirming the suitability of a model based on damage saturation effects. The results of *ab initio* simulation of the Young's modulus values were therefore fitted with Eq. (7), allowing the determination of the empirical value $\kappa = 4.46$, which is reasonably close to the reference value of $\kappa \approx \pi$ indicated for carbon nanotubes [17]. The numerical predictions of the Young's modulus variation derived from both the model and the *ab initio* simulations were then compared to experimental nanoindentation measurements, as reported in Fig. 6b. Both predictions are compatible with the experimental data within their respective uncertainties, although the theoretical predictions tend to systematically under-estimate the reduction in Young's modulus value. This tendency is confirmed by a previous experimental dataset from [26] (also included in Fig. 6b), which also falls below numerical predictions. This systematic underestimation by theoretical models can be reasonably attributed to the fact that in all of the above-mentioned approaches only the effect of induced isolated vacancies is modelled, while the possible effects of more complex defect aggregates are disregarded. Nonetheless, the compatibility between theoretical and experimental results is striking, thus indicating that (at least at the reported damage densities, i.e. below ~$10^{22}$ vacancies cm$^{-3}$) the isolated vacancy defect indeed plays a prominent role in the modification of the mechanical properties of diamond.

Finally, we verified the consistency of the QFM-based model with SPM measurements of the damage-induced surface swelling by feeding its predictions of the local variations of mass density (see Eq. (6) and Fig. 6a) and Young's modulus (see Eq. (7) and Fig. 6b) into the FEM simulation of the deformation of the ion-irradiated diamond region. The results are shown in Fig. 7a: the calculated displacement field in the implanted area and in the surrounding diamond crystal is visualized in colour scale, highlighting the localized deformation that reaches a maximum on the top surface in correspondence with the Bragg peak (see also Figs. 1 and 5). As shown in Fig. 7b, the calculated displacements agree very well with the measured surface swelling, with the exception of the region extending beyond the ion end of range depth at ~25 μm, where the theoretical predictions of the surface swelling significanty overestimate the experimental results. This is possibly due to the isotropic (instead of cubic) symmetry used in FEM simulations for both implanted and unimplanted diamond: such an approximation is more appropriate for the partially amorphized implanted material. The SPM measurements show non-zero surface swelling effects (~22.5 nm) at the leading edge of the implanted region, where the vacancy density is negligible, and this feature is correctly captured by FEM simulations. Since the Young's modulus variation used in the FEM simulations is derived from QFM calculations with a $\kappa$ interaction parameter derived from *ab initio* simulations, the agreement between FEM-calculated and experimentally measured surface displacements represents a significant confirmation of the consistency of obtained results.

FEM simulations based on the predictions of multi-scale modelling of the variation of the structural/mechanical properties in ion-implanted diamond can also provide reliable estimates of locally induced strains and stresses, including in the bulk of the implanted crystal, which is not accessible to direct measurements. This is essential when performing diamond microfabrication, since it was recently shown that diamond amorphization is a strain-driven process [48], i.e. it occurs in regions where a defined strain threshold (~16%) is exceeded. Therefore, such a strain (or equivalently, stress) value must be estimated with the highest accuracy. Here, as an example we report in Fig. 7c the Von Mises stress profiles along the direction of the incident ion beam, as calculated at three different positions along the $z$ axis, from the top surface ($z = 0$ μm) to the lower border of the implanted region ($z = -h = -100$ μm). As expected, all stress profiles display a pronounced peak at the end-of-range depth of the ions, where the defect density is largest, reaching values of up to 12 GPa at the sample surface. Moreover, a 3-fold decrease of the end-of-range stress value is observed at a depth of ~100 μm. This evidence provides significant insight into an open issue in the state of the art, i.e. the depth dependence of the amorphization threshold for diamond [23, 32, 33, 58-65]. The general (but so far undemonstrated) understanding is that such a threshold increases at increasing implantation depth due to higher internal pressures that do not allow relaxation to graphitic structures [32]. Here we show quantitatively that in fact the opposite is true, i.e. deeper implants induce smaller stresses (and strains). This is nevertheless consistent with experimental observations, i.e. higher defect densities are required at increasing depths to achieve the same amorphization-inducing strain values. The estimated 3-fold increase in amorphisation threshold for a 100 μm depth is consistent with experimentally observed variations in the literature [23, 32, 33, 58-65].

## 5. Conclusions

A systematic study of the variation of the structural and mechanical properties of defected diamond was carried out with cross-sectional Raman, SPM and nano-indentation measurements carried on MeV-ion-implanted samples. The experimental results display a very satisfactory agreement with complementary *ab initio* atomistic and mesoscale models, also when integrated in FEM simulations. Measurements show that a softening effect of up to 15% in diamond, i.e. the hardest known bulk material, can potentially be obtained in a controlled manner using ion irradiation and/or FIB techniques without incurring in the full amorphization/graphitization of the pristine crystal. Simulations predict that this effect can reach nearly 50% for higher damage densities (~$2.2\times10^{22}$ cm$^{-3}$), still lying below the graphitization threshold for diamond (~$2.8\times10^{22}$ cm$^{-3}$) [48]. The proposed approach opens the way to the possibility of achieving a three-dimensional tuning of the Young's modulus of diamond with a micrometric spatial resolution by adopting advanced microfabrication techniques, e.g. lateral irradiation configurations, multiple implantations, and/or appropriate masking processes. Furthermore, this work demonstrates the possibility of obtaining through FEM simulations an accurate mapping of the stress fields present in the ion-damaged material, allowing for a useful control during various microfabrication stages. Simulations show that deeper implants require a higher density of induced defects to promote amorphization, confirming both the strain-driven mechanism proposed in previous works [48] and the generally observed increase of the graphitization threshold for deeper ion implantations [23, 32, 33, 58-65].

## Acknowledgements

This work is supported by the following projects: "DiNaMo" (young researcher grant, project n. 157660) by Italian National Institute of Nuclear Physics; FIRB "Futuro in Ricerca 2010" (CUP code: D11J11000450001) funded by MIUR and "A.Di.N-Tech." (CUP code: D15E13000130003), "Linea 1A - ORTO11RRT5" projects funded by the University of Torino and "Compagnia di San Paolo". Nanofacility Piemonte is laboratory supported by the "Compagnia di San Paolo" foundation. N.M.P. is supported by the European Research Council (ERC StG Ideas 2011 n. 279985 BIHSNAM, ERC PoC 2013 n. 632277 KNOTOUGH, ERC PoC 2015 n. 693670 SILKENE), and by the EU under the FET

Graphene Flagship (WP 14 "Polymer nanocomposites" n. 696656). F.B. is supported by BIHSNAM.

**References**

[1]     M.Y. Liao, S. Hishita, E. Watanabe, S. Koizumi, Y. Koide. Suspended Single-Crystal Diamond Nanowires for High-Performance Nanoelectromechanical Switches, Advanced Materials 22 (2010) 5393-+.
[2]     M.K. Zalalutdinov, M.P. Ray, D.M. Photiadis, J.T. Robinson, J.W. Baldwin, J.E. Butler, T.I. Feygelson, B.B. Pate, B.H. Houston. Ultrathin Single Crystal Diamond Nanomechanical Dome Resonators, Nano Letters 11 (2011) 4304-4308.
[3]     M.J. Burek, N.P. de Leon, B.J. Shields, B.J.M. Hausmann, Y. Chu, Q. Quan, A.S. Zibrov, H. Park, M.D. Lukin, M. LonÄ•ar. Free-Standing Mechanical and Photonic Nanostructures in Single-Crystal Diamond, Nano Letters 12 (2012) 6084-6089.
[4]     Y. Tao, J.M. Boss, B.A. Moores, C.L. Degen. Single-crystal diamond nanomechanical resonators with quality factors exceeding one million, Nat Commun 5 (2014).
[5]     O. Auciello, J. Birrell, J.A. Carlisle, J.E. Gerbi, X.C. Xiao, B. Peng, H.D. Espinosa. Materials science and fabrication processes for a new MEMS technoloey based on ultrananocrystalline diamond thin films, J Phys-Condens Mat 16 (2004) R539-R552.
[6]     A. Gaidarzhy, M. Imboden, P. Mohanty, J. Rankin, B.W. Sheldon. High quality factor gigahertz frequencies in nanomechanical diamond resonators, Applied Physics Letters 91 (2007) 203503.
[7]     M.W. Doherty, N.B. Manson, P. Delaney, F. Jelezko, J. Wrachtrup, L.C.L. Hollenberg. The nitrogen-vacancy colour centre in diamond, Physics Reports 528 (2013) 1-45.
[8]     T. Mueller, C. Hepp, B. Pingault, E. Neu, S. Gsell, M. Schreck, H. Sternschulte, D. Steinmueller-Nethl, C. Becher, M. Atatuere. Optical signatures of silicon-vacancy spins in diamond, Nat Commun 5 (2013) 3328.
[9]     T.H. Taminiau, CramerJ, T. van der Sar, V.V. Dobrovitski, HansonR. Universal control and error correction in multi-qubit spin registers in diamond, Nat Nano 9 (2014) 171-176.
[10]    D.D. Awschalom, R. Epstein, R. Hanson. The diamond age of spintronics - Quantum electronic devices that harness the spins of electrons might one day enable room-temperature quantum computers - made of diamond, Sci Am 297 (2007) 84-+.
[11]    E.R. MacQuarrie, T.A. Gosavi, N.R. Jungwirth, S.A. Bhave, G.D. Fuchs. Mechanical Spin Control of Nitrogen-Vacancy Centers in Diamond, Phys Rev Lett 111 (2013) 227602.
[12]    B.J.M. Hausmann, B.J. Shields, Q. Quan, Y. Chu, N.P. de Leon, R. Evans, M.J. Burek, A.S. Zibrov, M. Markham, D.J. Twitchen, H. Park, M.D. Lukin, M. Lončar. Coupling of NV Centers to Photonic Crystal Nanobeams in Diamond, Nano Letters 13 (2013) 5791-5796.
[13]    P. Ovartchaiyapong, K.W. Lee, B.A. Myers, A.C.B. Jayich. Dynamic strain-mediated coupling of a single diamond spin to a mechanical resonator, Nature Communications 5 (2014) 5429.
[14]    G. Lopez-Polin, C. Gomez-Navarro, V. Parente, F. Guinea, M.I. Katsnelson, F. Perez-Murano, J. Gomez-Herrero. Increasing the elastic modulus of graphene by controlled defect creation, Nature Physics 11 (2015) 26-31.
[15]    A. Zandiatashbar, G.H. Lee, S.J. An, S. Lee, N. Mathew, M. Terrones, T. Hayashi, C.R. Picu, J. Hone, N. Koratkar. Effect of defects on the intrinsic strength and stiffness of graphene, Nature Communications 5 (2014) 3186.
[16]    M.B. Nardelli, B.I. Yakobson, J. Bernholc. Brittle and ductile behavior in carbon nanotubes, Phys Rev Lett 81 (1998) 4656-4659.
[17]    N.M. Pugno. Young's modulus reduction of defective nanotubes, Applied Physics Letters 90 (2007) 043106
[18]    B.A. Fairchild, P. Olivero, S. Rubanov, A.D. Greentree, F. WaIdermann, R.A. Taylor, I. Walmsley, J.M. Smith, S. Huntington, B.C. Gibson, D.N. Jamieson, S. Prawer. Fabrication of Ultrathin Single-Crystal Diamond Membranes, Adv Mater 20 (2008) 4793-+.
[19]    Y.C. Song, M.L. Lee. Room temperature electroluminescence from light-emitting diodes based on In0.5Ga0.5As/GaP self-assembled quantum dots, Applied Physics Letters 100 (2012) 251904

[20]   I. Aharonovich, J.C. Lee, A.P. Magyar, B.B. Buckley, C.G. Yale, D.D. Awschalom, E.L. Hu. Homoepitaxial growth of single crystal diamond membranes for quantum information processing, Adv Mater 24 (2012) OP54-59.
[21]   A.A. Martin, S. Randolph, A. Botman, M. Toth, I. Aharonovich. Maskless milling of diamond by a focused oxygen ion beam, Sci Rep-Uk 5 (2015) 8958.
[22]   J.P. Hadden, J.P. Harrison, A.C. Stanley-Clarke, L. Marseglia, Y.L.D. Ho, B.R. Patton, J.L. O'Brien, J.G. Rarity. Strongly enhanced photon collection from diamond defect centers under microfabricated integrated solid immersion lenses, Applied Physics Letters 97 (2010) 241901
[23]   J.F. Prins, T.E. Derry. Radiation defects and their annealing behaviour in ion-implanted diamonds, Nuclear Instruments & Methods in Physics Research Section B-Beam Interactions with Materials and Atoms 166 (2000) 364-373.
[24]   J.F. Prins, T.E. Derry, J.P.F. Sellschop. Volume Expansion of Diamond during Ion-Implantation, Phys Rev B 34 (1986) 8870-8874.
[25]   P. Olivero, F. Bosia, B.A. Fairchild, B.C. Gibson, A.D. Greentree, P. Spizzirri, S. Prawer. Splitting of photoluminescent emission from nitrogen-vacancy centers in diamond induced by ion-damage-induced stress, New J Phys 15 (2013).
[26]   R.A. Khmelnitsky, V.A. Dravin, A.A. Tal, M.I. Latushko, A.A. Khomich, A.V. Khomich, A.S. Trushin, A.A. Alekseev, S.A. Terentiev. Mechanical stresses and amorphization of ion-implanted diamond, Nucl Instrum Meth B 304 (2013) 5-10.
[27]   F. Bosia, N. Argiolas, M. Bazzan, B.A. Fairchild, A.D. Greentree, D.W.M. Lau, P. Olivero, F. Picollo, S. Rubanov, S. Prawer. Direct measurement and modelling of internal strains in ion-implanted diamond, Journal of Physics-Condensed Matter 25 (2013) 385403.
[28]   C.A. Klein, G.F. Cardinale. Young's modulus and Poisson's ratio of CVD diamond, Diam Relat Mater 2 (1993) 918-923.
[29]   P. Djemia, A. Tallaire, J. Achard, F. Silva, A. Gicquel. Elastic properties of single crystal diamond made by CVD, Diam Relat Mater 16 (2007) 962-965.
[30]   Y.X. Wei, R.J. Wang, W.H. Wang. Soft phonons and phase transition in amorphous carbon, Phys Rev B 72 (2005) 012203.
[31]   J.F. Ziegler, M.D. Ziegler, J.P. Biersack. SRIM - The stopping and range of ions in matter (2010), Nucl Instrum Meth B 268 (2010) 1818-1823.
[32]   J.O. Orwa, K.W. Nugent, D.N. Jamieson, S. Prawer. Raman investigation of damage caused by deep ion implantation in diamond, Phys Rev B 62 (2000) 5461-5472.
[33]   R. Kalish, A. Reznik, K.W. Nugent, S. Prawer. The nature of damage in ion-implanted and annealed diamond, Nucl Instrum Meth B 148 (1999) 626-633.
[34]   W. Wu, S. Fahy. Molecular-Dynamics Study of Single-Atom Radiation-Damage in Diamond, Phys Rev B 49 (1994) 3030-3035.
[35]   D.N. Jamieson, S. Prawer, K.W. Nugent, S.P. Dooley. Cross-sectional Raman microscopy of MeV implanted diamond, Nucl Instrum Meth B 106 (1995) 641-645.
[36]   F. Bosia, N. Argiolas, M. Bazzan, P. Olivero, F. Picollo, A. Sordini, M. Vannoni, E. Vittone. Modification of the structure of diamond with MeV ion implantation, Diam Relat Mater 20 (2011) 774-778.
[37]   W.C. Oliver, G.M. Pharr. Measurement of hardness and elastic modulus by instrumented indentation: Advances in understanding and refinements to methodology, Journal of Materials Research 19 (2004) 3-20.
[38]   R. Dovesi, R. Orlando, A. Erba, C.M. Zicovich-Wilson, B. Civalleri, S. Casassa, L. Maschio, M. Ferrabone, M. De La Pierre, P. D'Arco, Y. Noel, M. Causa, M. Rerat, B. Kirtman. CRYSTAL14: A Program for the Ab Initio Investigation of Crystalline Solids, Int J Quantum Chem 114 (2014) 1287-1317.
[39]   V.R.S. R. Dovesi, C. Roetti, R.Orlando, C. M. Zicovich-Wilson,C.Pascale, B. Civalleri,K.Doll, N.M. Harrison,I.J.Bush, Ph. D'Arco, M. Llunel, M.Causà,Y.Noel. CRYSTAL14 Manual. http://www.crystal.unito.it/Manuals/crystal14.pdf.
[40]   W.F. Perger, J. Criswell, B. Civalleri, R. Dovesi. Ab-initio calculation of elastic constants of crystalline systems with the CRYSTAL code, Comput Phys Commun 180 (2009) 1753-1759.

[41] R. Orlando, R. Dovesi, P. Azavant, N.M. Harrison, V.R. Saunders. A Super-Cell Approach for the Study of Localized Defects in Solids - Carbon Substitution in Bulk Silicon, J Phys-Condens Mat 6 (1994) 8573-8583.
[42] A. Erba, A. Mahmoud, R. Orlando, R. Dovesi. Elastic properties of six silicate garnet end members from accurate ab initio simulations, Phys Chem Miner 41 (2014) 151-160.
[43] A. Erba, A. Mahmoud, D. Belmonte, R. Dovesi. High pressure elastic properties of minerals from ab initio simulations: The case of pyrope, grossular and andradite silicate garnets, Journal of Chemical Physics 140 (2014).
[44] C.T. Lee, W.T. Yang, R.G. Parr. Development of the Colle-Salvetti Correlation-Energy Formula into a Functional of the Electron-Density, Phys Rev B 37 (1988) 785-789.
[45] A.E.a.E. Albanese. Elastic, Piezoelectric and Photoelastic Tensors Tutorial. www.theochem.unito.it/crystal_tuto/mssc2013_cd/tutorials/index.html.
[46] J.A. van Wyk, O.D. Tucker, M.E. Newton, J.M. Baker, G.S. Woods, P. Spear. Magnetic-resonance measurements on the $^{5}$$\mathit{A}_{2}$ excited state of the neutral vacancy in diamond, Phys Rev B 52 (1995) 12657-12667.
[47] J.E. Lowther, A. Mainwood. A perturbed vacancy model for the R1 EPR centre in diamond, Journal of Physics: Condensed Matter 6 (1994) 6721.
[48] B.A. Fairchild, S. Rubanov, D.W.M. Lau, M. Robinson, I. Suarez-Martinez, N. Marks, A.D. Greentree, D. McCulloch, S. Prawer. Mechanism for the Amorphisation of Diamond, Adv Mater 24 (2012) 2024-2029.
[49] F. Bosia, S. Calusi, L. Giuntini, S. Lagomarsino, A. Lo Giudice, M. Massi, P. Olivero, F. Picollo, S. Sciortino, A. Sordini, M. Vannoni, E. Vittone. Finite element analysis of ion-implanted diamond surface swelling, Nucl Instrum Meth B 268 (2010) 2991-2995.
[50] N.M. Pugno, R.S. Ruoff. Quantized fracture mechanics, Philos Mag 84 (2004) 2829-2845.
[51] Y.X. Wei, R.J. Wang, W.H. Wang. Soft phonons and phase transition in amorphous carbon, Phys Rev B 72 (2005).
[52] R. Kalish, A. Reznik, S. Prawer, D. Saada, J. Adler. Ion-implantation-induced defects in diamond and their annealing: Experiment and simulation, Phys Status Solidi A 174 (1999) 83-99.
[53] S. Prawer, K.W. Nugent, D.N. Jamieson. The Raman spectrum of amorphous diamond, Diam Relat Mater 7 (1998) 106-110.
[54] T.V. Kononenko, A.A. Khomich, V.I. Konov. Peculiarities of laser-induced material transformation inside diamond bulk, Diam Relat Mater 37 (2013) 50-54.
[55] A.C. Ferrari, B. Kleinsorge, N.A. Morrison, A. Hart, V. Stolojan, J. Robertson. Stress reduction and bond stability during thermal annealing of tetrahedral amorphous carbon, J Appl Phys 85 (1999) 7191-7197.
[56] S.M. Pimenov, A.A. Khomich, I.I. Vlasov, E.V. Zavedeev, A.V. Khomich, B. Neuenschwander, B. Jaggi, V. Romano. Metastable carbon allotropes in picosecond-laser-modified diamond, Appl Phys a-Mater 116 (2014) 545-554.
[57] A.F. Bower. Applied mechanics of solids, CRC Press, Boca Raton, 2010.
[58] P. Olivero, S. Rubanov, P. Reichart, B.C. Gibson, S.T. Huntington, J.R. Rabeau, A.D. Greentree, J. Salzman, D. Moore, D.N. Jamieson, S. Prawer. Characterization of three-dimensional microstructures in single-crystal diamond, Diam Relat Mater 15 (2006) 1614-1621.
[59] A.A. Gippius, R.A. Khmelnitskiy, V.A. Dravin, S.D. Tkachenko. Formation and characterization of graphitized layers in ion-implanted diamond, Diam Relat Mater 8 (1999) 1631-1634.
[60] D.P. Hickey, K.S. Jones, R.G. Elliman. Amorphization and graphitization of single-crystal diamond - A transmission electron microscopy study, Diam Relat Mater 18 (2009) 1353-1359.
[61] J.F. Prins. C+-damaged diamond: electrical measurements after rapid thermal annealing to 500 degrees C, Diam Relat Mater 10 (2001) 463-468.
[62] P.F. Lai, S. Prawer, L.A. Bursill. Recovery of diamond after irradiation at high energy and annealing, Diam Relat Mater 10 (2001) 82-86.
[63] R. Brunetto, G.A. Baratta, G. Strazzulla. Raman spectroscopy of ion irradiated diamond, Journal of Applied Physics 96 (2004) 380-386.
[64] W.R. McKenzie, M.Z. Quadir, M.H. Gass, P.R. Munroe. Focused Ion beam implantation of diamond, Diam Relat Mater 20 (2011) 1125-1128.

[65]	V.S. Drumm, A.D.C. Alves, B.A. Fairchild, K. Ganesan, J.C. McCallum, D.N. Jamieson, S. Prawer, S. Rubanov, R. Kalish, L.C. Feldman. Surface damage on diamond membranes fabricated by ion implantation and lift-off, Applied Physics Letters 98 (2011) 231904.

## List of Figure captions

Fig. 1: Schematic representation (not to scale) of the experimental configuration: MeV ion implantation (red arrow) was performed on a lateral polished surface. The corresponding damage profile derived from Eq. (2) is reported in the inset graph on the right. Scanning nano-indentation and SPM measurements were carried out on the upper surface of the sample.

Fig. 2: a) Scanning Probe Microscopy map of the region of interest. The line labeled as "1" shows a typical scan along which nanoindentation measurements were performed. The end-of-range peak (highlighted by a dotted line) is visible at about 25 μm from the surface. b) Force-displacement nanoindentation curves for implanted (continuous blue line) and unimplanted (dashed black line) regions of the diamond sample.

Fig. 3: Defected diamond unit cells considered in ab initio simulations, ranging from 128- to 8-atom systems with a single vacancy at their body centres.

Fig. 4: a) Micro-Raman spectra acquired across the highly damaged region; the spectra are displaced along the vertical axis for sake of readability; the blue spectrum corresponds to the Bragg peak region at the end of ion range; b) corresponding values of the shift of the first-order Raman peak measured at different positions across the sample depth; the highlighted datapoints correspond to spectra of the same color in a).

Fig. 5: SPM topography (blue circular dots, scale on the left) and Young's modulus (red square dots, scale on the right) depth profiles of the implanted area acquired along the same linear scan in correspondence with the end-of-range Bragg peak.

Fig. 6: a) mass density percentage reduction obtained as a function of vacancy density from ab initio (blue diamond and red square dots for HF and B3LYP Hamiltonians, respectively) and mesoscale (continuous black line) simulations; b) Young's modulus percentage reduction as a function of vacancy density obtained in nanoindentation experiments (black triangular dots), from ab initio simulations using HF and B3LYP hamiltonians (blue diamond and red square dots, respectively), from Quantum Fracture Mechanics calculations (continuous black line) and from experimental data in the literature [26] (green circular dots).

Fig. 7: a) FEM-simulated displacements (in colour scale) in a laterally implanted diamond sample, on an area of height h = 100 μm. b) SPM topography and FEM simulations representing the swelling effect over the whole ion penetration depth (left scale) and SRIM vacancy density profile (right scale). c) Calculated Von Mises stress variation in the y direction at three different depths from the sample surface: z = 0 μm, −50 μm and −100 μm.

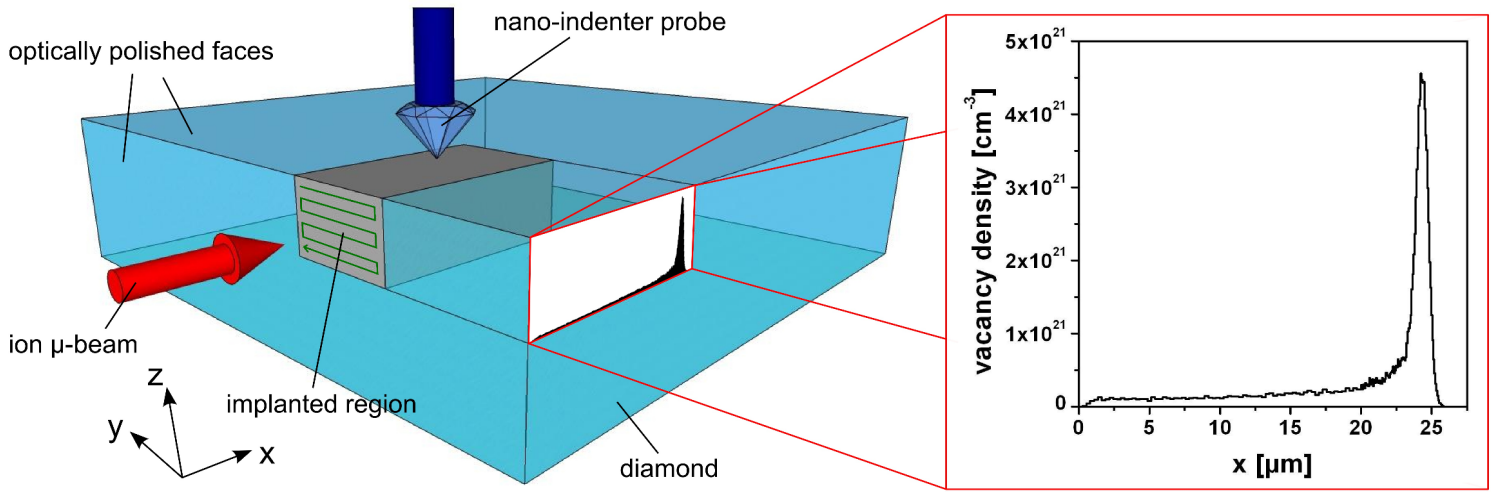

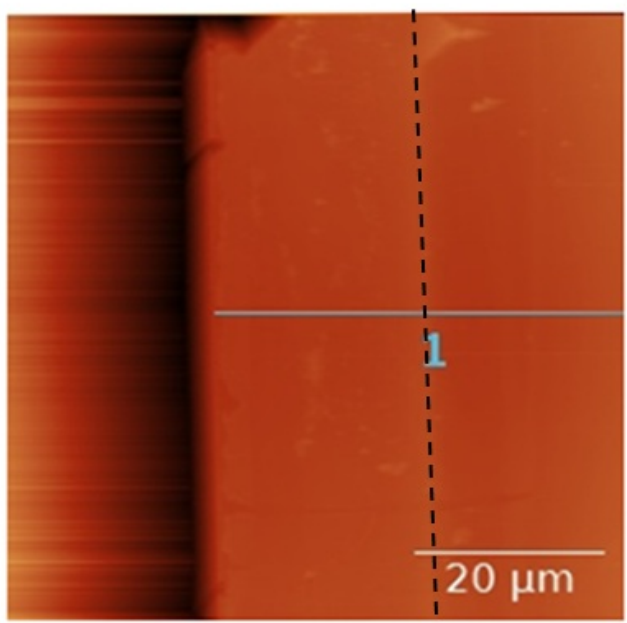 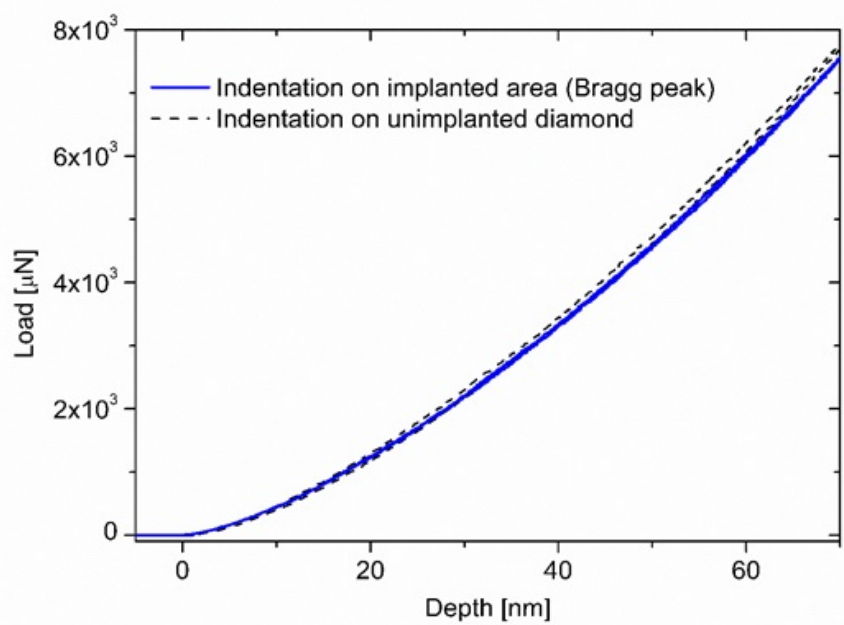

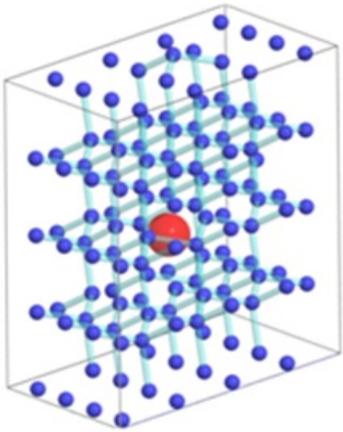 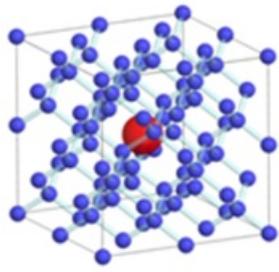 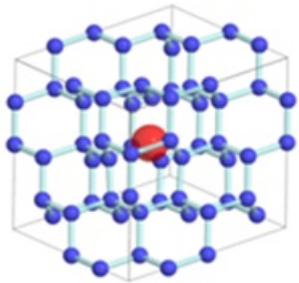 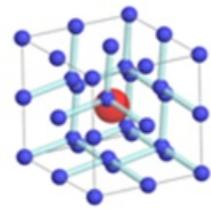 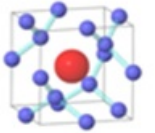

128　　　　　　64　　　　　　32　　　　　　16　　　　　　8

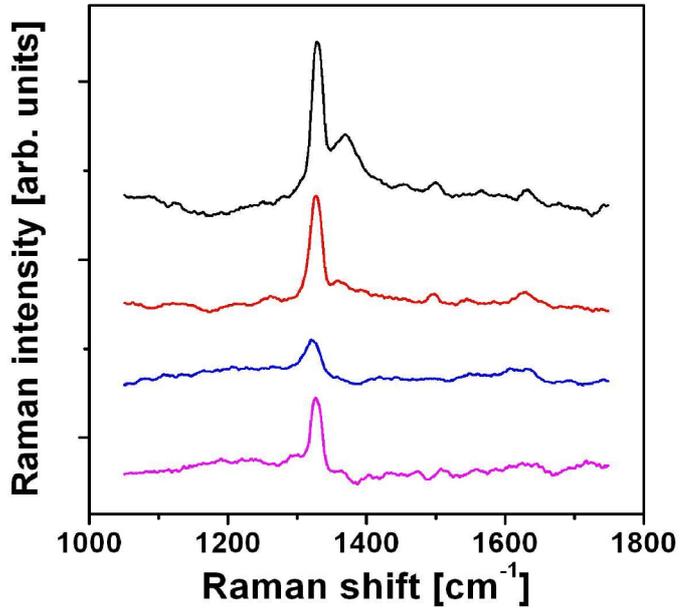 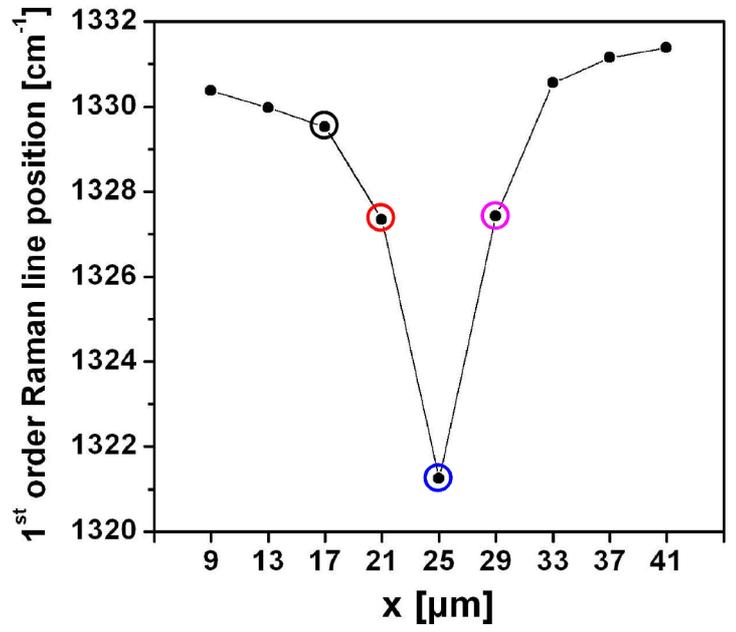

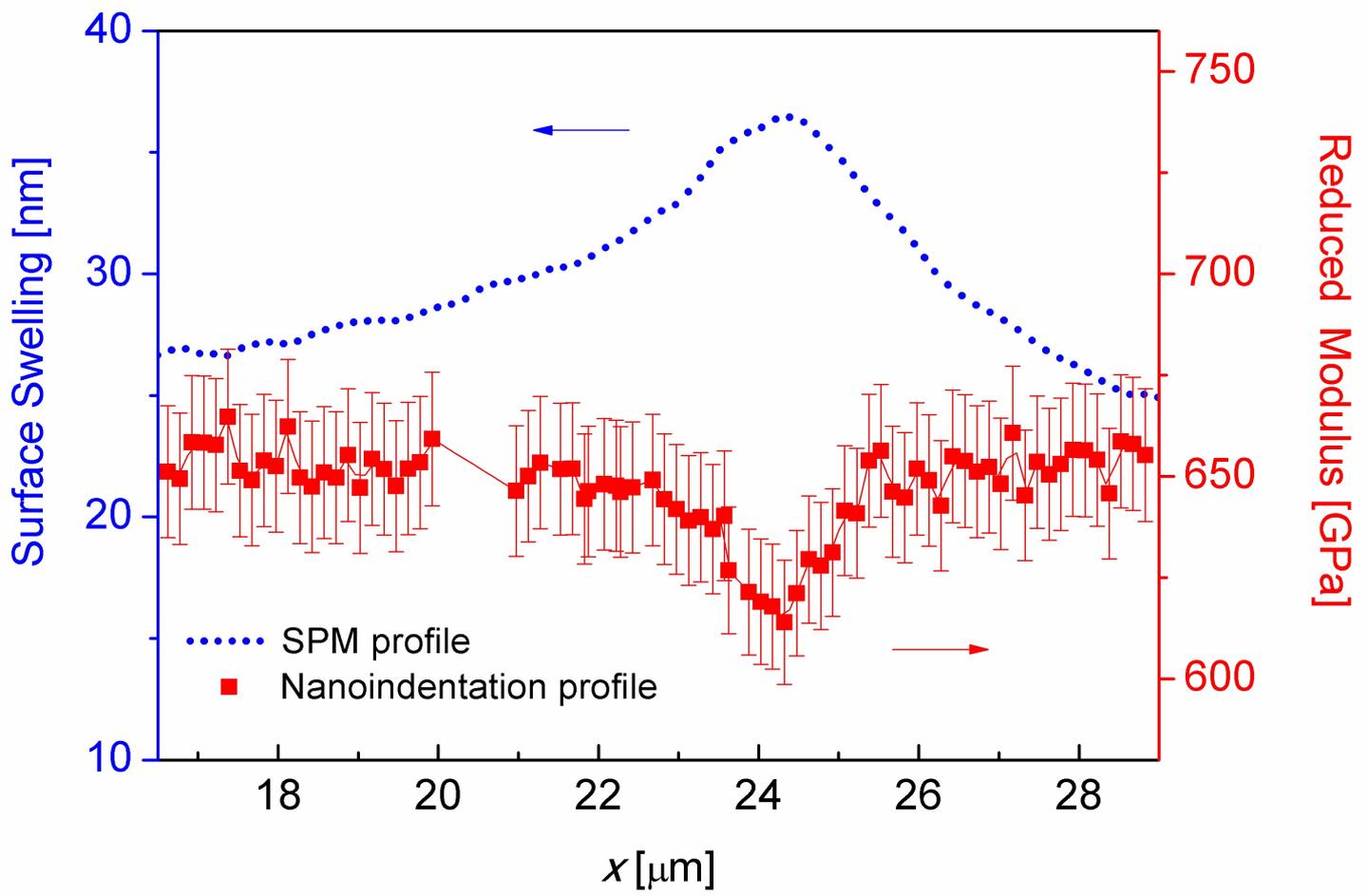

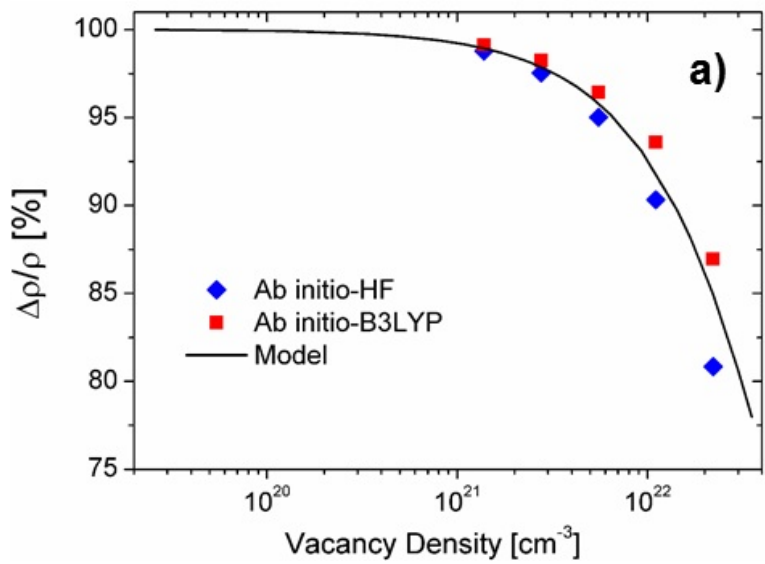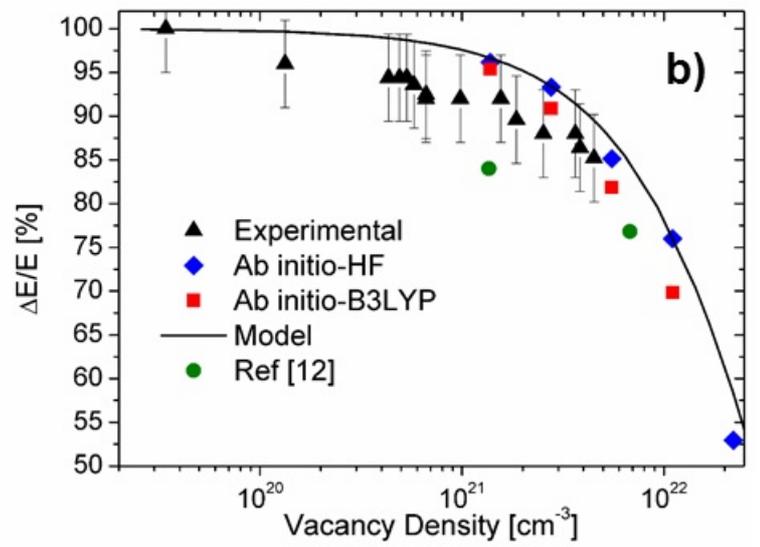

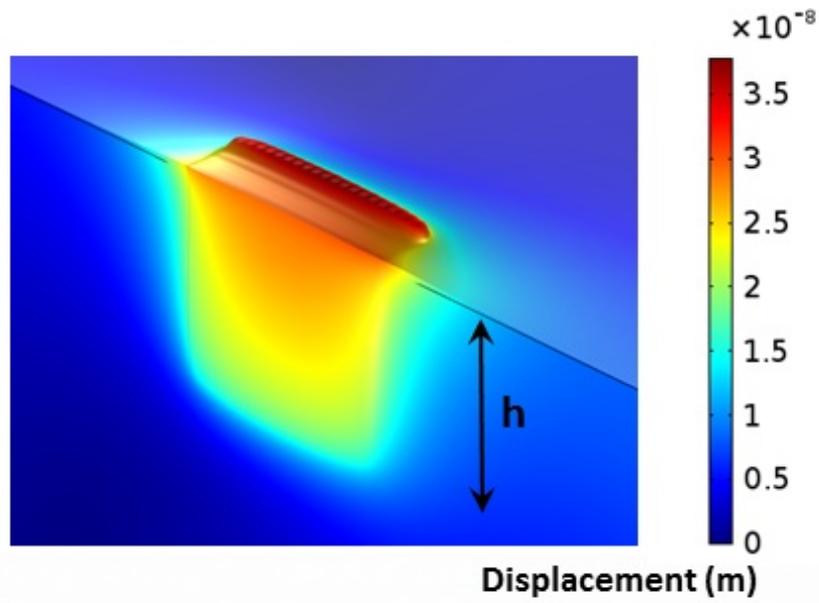
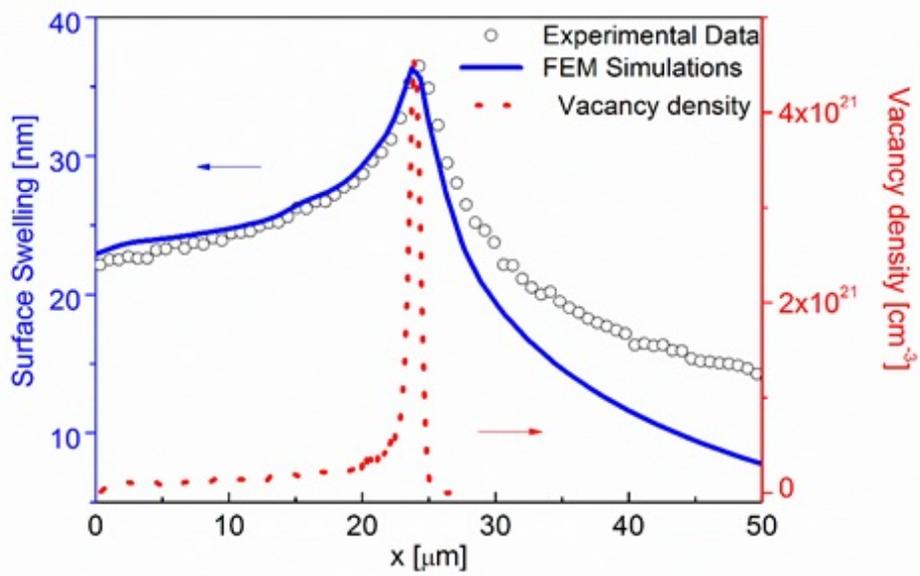
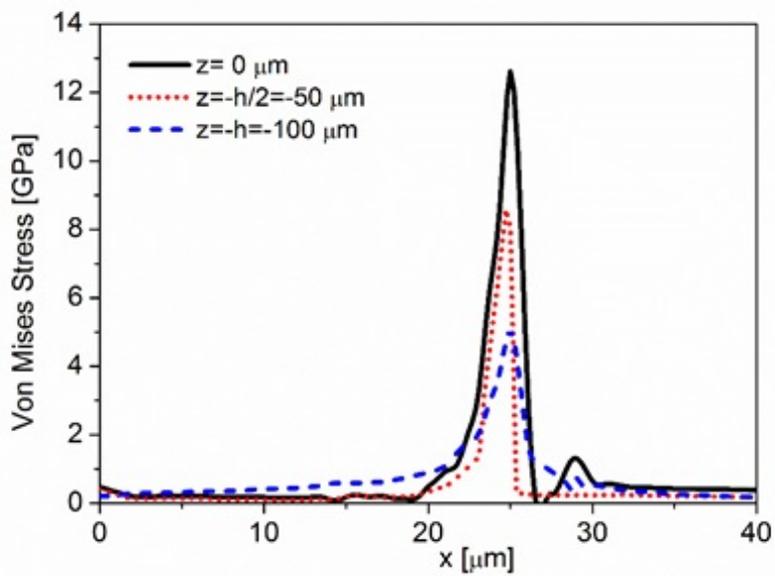